\documentclass[pdflatex,sn-mathphys]{sn-jnl}

\jyear{2021}%

\theoremstyle{thmstyleone}%
\theoremstyle{thmstyletwo}%

\theoremstyle{thmstylethree}%

\begin{document}

\title[Monu yadav and Laxminarayan Das]{Analyze the SATCON Algorithm's Capability to Estimate Tropical Storm Intensity across the West Pacific Basin}


\author[1]{\fnm{Monu} \sur{Yadav}}\email{yadavm012@gmail.com}

\author*[1]{\fnm{Laxminarayan} \sur{Das}}\email{lndas@dce.ac.in}


\affil*[1]{\orgdiv{Department of Applied Mathematics}, \orgname{Delhi Technological University}, \orgaddress{ \city{New Delhi}, \country{India}}}




\abstract{A group of algorithms for estimating the current intensity (CI) of tropical cyclones (TCs), which use infrared and microwave sensor-based images as the input of the algorithm because it is more skilled than each algorithm separately, are used to create a technique to estimate the TC intensity which is known as SATCON . In the current study, an effort was undertaken to assess how well the SATCON approach performed for estimating TC intensity throughout the west pacific basin from year 2017 to 2021. To do this, 26 TCs over the west pacific basin were analysed using the SATCON-based technique, and the estimates were compared to the best track parameters provided by the Regional Specialized Meteorological Centre (RSMC), Tokyo. The maximum sustained surface winds (Vmax) and estimated central pressures (ECP) for various ``T" numbers and types of storm throughout the entire year as well as during the pre-monsoon (March-July) and post-monsoon (July-February) seasons have been compared. When compared to weaker and very strong TCs, the ability of the SATCON algorithm to estimate intensity is determined to be rather excellent for mid-range TCs. We demonstrate that SATCON is more effective in the post-monsoon across the west pacific basin than in the pre-monsoon by comparing the algorithm results.}




\maketitle

\section{Introduction}\label{sec1}
Tropical cyclone (TC) observation by meteorological satellites has largely reduced the challenge of detection. A constellation of geostationary (GEO) and polar-orbiting platforms regularly scans the tropics, and sensors with better spatial and spectrum sampling are used. Numerous types of multispectral photography can be used to qualitatively track and record the location, genesis, occurrence, and dissipation of TCs. Estimating the present TC intensity from space-based platforms is a little more complicated. It is possible to perform subjective analysis of TC cloud patterns using infrared (IR) images employing trained analysts and empirically supported guidelines. In order to analyse the CI and anchor TC intensity catalogues (or ``best tracks") in the absence of in situ intensity observations, operational TC centres have depended extensively on the time-tested Dvorak technique \cite{bib8,bib9} for many years. As demonstrated by crowdsourcing techniques, even inexperienced analysts can fairly accurately estimate the CI \cite{bib10}. The inherent subjectivity in the interpretation of the images and restrictions on the capacity to detect structured convective structure beneath the normally massive and dense TC cirrus canopy, however, pose difficulties to IR-based cloud pattern recognition techniques \cite{bib11,bib12}. Techniques that make use of cloud-penetrating microwave (MW) sensors \cite{bib13,bib14,bib15,bib16} can be helpful in this area, but they also have drawbacks.

Obtaining accurate CI estimations is crucial for various reasons: The operational TC forecast process begins with the CI; it is one of the key input variables required to initialise both dynamical and statistical TC forecast models; and it is crucial for understanding TC climatologies and trends to have precise best-track intensities \cite{bib1}. Forecasters (or best-track analysts) frequently struggle with the issue of competing satellite-based CI estimations with significant spread/uncertainty. Taken as a solution, a common conservative strategy is to average the estimations (simple consensus). A ``smarter" consensus procedure, depending on the situational performance of each consensus attribute, is preferred since it further minimises the CI estimate uncertainty.

Multiple satellite fitted sensor respond data based observation techniques are combined into an ensemble model known as SATCON created by Cooperative Institute for Meteorological Satellite Studies (CIMSS). Below are basic explanations of SATCON's methodology.
 
 Each attribute has situational strengths and weaknesses, which are represented by their separate intensity estimation error distributions, from which the individual attribute weights utilised in the SATCON process of finding a CI are created. Therefore, each attribute's performance behaviour can be categorised into situational bins. For example, using the IR images of the scene type, the ADT technique \cite{bib17} estimates the intensity of TC. When the eye scene is clear, it provides the best estimation; nevertheless, if it is not clear, the estimation is poor or the outcome may be compromised. To best weight all of the available intensity estimates into a single, superior consensus estimate, SATCON uses this situational information. The two TC intensity measurements, MSLP and MSW, each have distinct performance traits, leading to various SATCON weighting algorithms for each metric.
 
 Sharing information between sensors is another aspect of the SATCON process. Each SATCON attribute contains distinctive parametric data that the other coinciding attribute might use to evaluate the situational bins and conceivably modify the intensity estimates.For example, when an eye is observable in the IR, then ADT generates estimations of TC eye size \cite{bib18}.
 
 The Automated Tropical Cyclone Forecasting system (ATCF; \cite{bib19}) can be utilised by operational TC centres to provide additional sources of input to the SATCON process. These sources can include storm motion and the environmental pressure used in the pressure $>$ wind. For storms that significantly differ from an average TC motion of roughly 11 kts ($1 kts \approx 0.51 ms ^{-1} $), the methodology from \cite{bib20} can be used to make minor modifications to the final predicted MSW values.
 
 In the current work, the authors made an effort to evaluate the SATCON algorithm's performance in estimating TCs intensity throughout the west pacific basin by contrasting it with the parameters supplied by RSMC Tokyo. Section \ref{sec2} describes the data and technique used based on the statistics of 2017-2021. Section \ref{sec3} discusses the findings, while section \ref{sec4} enumerates the study's conclusions.
\section{Data and Methodology} \label{sec2}
We verified the ``SATCON" output data by comparing it to the data provided by RSMC, Tokyo of TC intensity for all TCs between 2017 and 2021 over the West Pacific (particularly taking into account these storms that affect Japan). The SATCON algorithm's data collects from UW-CIMSS \url{(http://tropic.ssec.wisc.edu/misc/satcon)} and RSMC provided data collects from the RSMC, Tokyo \url{(jma.go.jp/jma/jma-eng/jma-center/rsmc-hp-pub-eg/RSMC_HP.htm)}, to determine the optimal track parameters for TCs. The number of TC cases included in the study is listed in Table \ref{tab1} \cite{bib7,bib6,bib5,bib4,bib3}. 26 TCs are therefore investigated in the current study.

\begin{table}[h]
    \centering
    \caption{In this study, following TCs were considered over the West Pacific Basin}
    \begin{tabular}{c c c c c c}
    	 
    \hline
        SI. No. & Cyclone Name & Season & Date & Maximum Wind  \\
        &&&& Speed (knots) \\
        \hline
        2021 \\
         1& Surigae & Pre-Monsoon & 12-30 Apr. & 120 kts \\
         2 & IN-FA & Pre-Monsoon & 15-30 Jul. & 85 kts \\
         3 & Chanthu & Post-Monsoon & 05-20 Sept. & 115 kts\\
         4 & Rai & Post-Monsoon & 11-21 Dec. & 105 kts\\
         2020 \\
         1 & Vongfone & Pre-Monsoon & 08-18 May & 85 kts \\
         2 & Maysak & Post-Monsoon & 27 Aug. - 07 Sept. & 95 kts\\
         3 & Haishen & Post-Monsoon & 30 Aug. - 10 Sept. & 105 kts \\
         4 & Goni & Post-Monsoon & 26 Oct. - 06 Nov. & 115 kts\\
         5 & Molave & Post-Monsoon & 22-29 Oct. & 90 kts \\
         2019 \\
         1 & Nari & Pre-Monsoon & 24-28 Jul.&35 kts \\
         2 & Danas & Pre-Monsoon & 14-23 Jul. & 45 kts \\
         3 & Lekima & Post-Monsoon &02-15 Aug.&105 kts \\
         4 & Wutip & Post-Monsoon & 08 Feb. - 02 Mar. & 105 kts\\
         5 & Hagibis & Post-Monsoon & 04-14 Oct. &105 kts \\
         6 & Halong & Post-Monsoon & 01-10 Nov. &115 kts\\
         2018 \\
         1 & Jelawat & Pre-Monsoon & 24 Mar. - 01 Apr.&105 kts\\
         2 & Prapiroon & Pre-Monsoon &28 Jun. - 05 Jul.& 65 kts \\
         3 & Maria & Pre-Monsoon & 03-13 Jul. & 105 kts \\
         4 & Shanshan & Post-Monsoon &02-11 Aug. & 70 kts \\
         5 & Trami & Post-Monsoon &20 Sept. - 03 Oct.&105 kts \\
         6 & Kong-Rey & Post-Monsoon &28 Sept. - 07 Oct. & 115 kts \\
         2017 \\
         1 & Noru & Pre-Monsoon & 19 July-12 Aug. & 95 kts \\
         2 & Talim & Post-Monsoon & 08-22 Sept. & 95 kts\\
         3 & Sanvu & Post-Monsoon & 26 Aug. - 06 Sept. & 80 kts \\
         4 & Lan & Post-Monsoon & 15-23 Oct. & 100 kts \\
         5 & Hato & Post-Monsoon & 19-24 Aug. & 75 kts \\
         \hline
    \end{tabular}
   
    \label{tab1}
\end{table}

Multiple satellite fitted sensor respond data based observation techniques are combined into an ensemble model known as SATCON. Which is the studies of Geostationary-based Advanced Dvorak Technique and the Passive Microwave signal based advance sounding and imaging unit designed by the CIMSS. It provides a consensus intensity estimatation of TCs across all the basins. It makes use of a statistically determined weighting system that maximises (minimises) to be evaluated consensus intensity for a variety of TC structures (weaknesses). The intensity computation is built from a series of formulae dependent on the number of attribute available, and the SATCON weights are proportional to the RMSE attribute values for the selected scenarios.

The three-part equation for SATCON \cite{bib1} is

\begin{equation*}
    SATCON = \frac{W_{1}W_{2}(W_{1} + W_{2})E_{3} +W_{1}W_{3}(W_{1} + W_{3})E_{2} + W_{2}W_{3}(W_{3} + W_{2})E_{1} }{W_{1}W_{2}(W_{1} + W_{2}) + W_{1}W_{3}(W_{1} + W_{3}) + W_{3}W_{2}(W_{3} + W_{2})}
\end{equation*}
where $E_{n}$ is the attribute n's intensity estimations and $W_{n}$ is the attribute n's weight (RMSE). The weights of attributes 1, 2, and 3 are $W_{1}$, $W_{2}$, and $W_{3}$, and the intensity estimations of attributes 1, 2, and 3 are $E_{1}$, $E_{2}$, and $E_{3}$.

The situational RMSE values for each of the attributes used to calculate the intensity estimate are known as attribute weights. The SATCON weighting structure's composition is intended to give more weight to a situational dependent attribute with the highest efficiency (among the available attributes). For instance, the equation above shows how higher RMSEs (weights) of $E_{1}$ and $E_{2}$ are added to $E_{3}$. Thus adding greater weight to the specific estimation $E_{3}$, if $E_{3}$ is the best-performing attribute in a given context. For those more uncertain estimates, less weight (relatively smaller RMSEs) is to be alloted ($E_{1}$ and $E_{2}$) \cite{bib1}.

One of the finest methods for estimating the TC intensity over the Atlantic and North Indian Oceans is the SATCON \cite{bib22}.

To evaluate the accuracy of the intensity forecasting and the effectiveness of the CIMSS-SATCON algorithm, 26 TCs are used to validate the method (table \ref{tab1}). Comparison of RSMC, Tokyo \url{(jma.go.jp/jma/jma-eng/jma-center/rsmc-hp-pub-eg/RSMC_HP.htm)} provided intensity estimation data with SATCON intensity estimates.

Between the estimation of estimated central pressure (ECP) and Vmax based on RSMC, Tokyo provided data, and SATCON calculation, various variables, which are root mean square difference (RMSD), actual mean difference (bias), and mean absolute difference (MAD), are determined. These variables are estimated for the various stages of a TC's ``T" number, as specified in the RSMC, Tokyo-provided intensity data, inside each three-hourly observation that is at 00, 03, 06, 09, 12, 15, and 21 UTC throughout the whole time period of a TC. The mean MAD, RMSD, and bias of intensity estimations across the West Pacific basin are estimated for various ``T" numbers during the different seasons as well as for the entire year based on all TCs taken into account. The student's t-test is used to determine whether there are any significant differences between the mean values over the West Pacific basin during the pre- and post-monsoon seasons.

When compared to the information provided by RSMC, Tokyo, the capability of SATCON has also been evaluated for various stages of TCs. Table \ref{tab2} displays the various TC stages used in RSMC, Tokyo.
\begin{table}[h]
	\centering
	\begin{tabular}{c c   }
	\hline
	Stage & Maximum Sustained Wind (knots; kts) \\
	\hline
	Tropical Storm & 34-48 kts \\
	Severe Tropical Storm & 48-64 kts \\
	Typhoon & 64-85 kts \\
	Very Strong Typhoon & 85-105 kts \\
	Violent Typhoon & 105-130 kts \\ 
		\hline
	\end{tabular}
	\caption{Different stage of TCs with maximum sustained wind used in RSMC, Tokyo}
	\label{tab2}
\end{table}
\section{Results and discussion}\label{sec3}

\subsection{Over the period of entire year, the capability of the ``SATCON" algorithm over Japan (West Pacific)}
\subsubsection{Capability of the SATCON algorithm for various ``T" number stages}

\begin{sidewaystable}
\sidewaystablefn%
\begin{center}
\begin{minipage}{\textheight}
\caption{Calculated different parameters (in terms of MSW and MSLP) based on SATCON and RSMC, Tokyo provided data for TCs during the year 2017-2021}\label{tab3}
\begin{tabular*}{\textheight}{@{\extracolsep{\fill}}lccccccccc@{\extracolsep{\fill}}}
\toprule%
Best &&Best &Best &&&&Mean&\\
 track & Total no. & track	&  track  & SATCON  & SATCON & BIAS  &   absolute & RMSD \\
 T No. & of cases & intensity  & intensity & intensity  & intensity  & (A-B) & difference &  \\
 && range&(MSW) (A\footnotemark[1])&range&MSW (B\footnotemark[1])&&& \\
\midrule
2.0&94&30-40&35&40-67&50.75&-15.75&15.4&17.98 \\
2.5&99&40-50&42.68&40-70&52.97&-10.29&10.81&13.57\\
3.0&97&50-60&52.89&45-74&57.5&-4.61&5.13&12.18\\
3.5&166&60-70&62.65&46-92&82.1&-5.56&7.71&13.03\\
4.0&114&70-75&71.84&62-98&76.11&-4.27&10.08&12.83\\
5.0&123&75-80&80&69-100&83.47&-3.47&8.37&10.09\\
5.5&133&80-95&86.85&73-114&88.36&-1.51&6.93&11.72\\
6.0&98&95-105&97.5&90-122&105.04&-7.54&11.05&10.38\\
6.5&43&105-115&105.7&112-138&125.15&-19.45&20.18&21.76\\
7.0&16&115-125&115.93&122-144&136.47&-20.57&21.06&22.85\\
\botrule
 MSLP (hPa)\\
\toprule
Best &&Best &Best &&&&Mean&\\
 track & Total no. & track	&  track  & SATCON  & SATCON & BIAS  &   absolute & RMSD \\
 T No. & of cases & intensity & intensity & intensity  & intensity  & (A-B) & difference &  \\
 && range&(MSLP) (A\footnotemark[1])&range&MSLP (B\footnotemark[1])&&& \\
 2.0&21&996-1003&1005.85&984-1004&1002.03&3.82&4.92&6.28 \\
 2.5&35&983-1002&972.54&981-1004&970.59&1.95&4.01&6.83\\
 3.0&158&983-996&971.86&980-1000&969.5&2.36&5.58&7.09\\
 3.5&116&978-991&970.58&967-999&966.34&4.24&7.16&9.43\\
 4.0&104&975-988&964.67&957-994&959.5&5.17&9.78&11.13\\
 5.0&82&964-982&988.18&950-980&981.43&6.75&8.16&9.72\\
 5.5&45&955-964&962.45&939-976&956.38&6.07&9.09&10.73\\
 6.0&36&946-956&958.57&934-970&955.36&3.21&7.89&8.29\\
 6.5&22&932-946&898.11&922-940&891.69&6.42&8.95&8.47\\
 7.0&17&922-930&865.35&920-935&867.83&-2.48&6.72&7.43\\
 \midrule
\end{tabular*}

\end{minipage}
\end{center}
\end{sidewaystable}

Table \ref{tab3} compares the capability of SATCON TC MSW and MSLP calculation to intensity estimation data provided by RSMC for TCs across the west Pacific basin during 2017-2021. The bias progressively declines as the ``T" number rises, but it gradually rises after T5.5, being roughly 16-10 knots (kts) for T2.0-T2.5, approximately 6-4 kts for T3.0-T5.0, and about 2 kts for T5.5. Due to the small sample size, the results for T6.0-T7.0 bias increasing with increasing in T number from roughly 8 to 21 kts may not be indicative. According to the student's t-test for the ``T" numbers T2.0-T5.5 and T6.0-T7.0, the difference is significant with a $99\%$ level of confidence.

After the T3.5, the MAD is approximately 10-14 kts, and the MAD is approximately 12-16 kts for T2.0-T.3.0 . Due to the small sample size, the higher MAD value in the T6.5 range could not be indicative. The MAD values for T5.0 and above across the west Pacific (7-10 kts) are consistent with \cite{bib1,bib2} observations.

As a consequence, the intensity is estimated to be overestimated (negative bias) by approximately 2 hPa for T7.0, approximately 2-5 hpa for T2.0-T3.5 and approximately 5-7 hPa for more than T3.5. For the range of T2.0-T6.5, the underestimate is statistically significant at a $99\%$ level of confidence. For T2.0-T2.5, the MAD is approximately 5 hPa, and for T3.0-T7.0, it is approximately 5-10 hPa. For T2.0-T3.0, the RMSD is about 6-7 hPa, and for T3.5-T7.0, it is approximately 8-11 hPa.
\subsubsection{Results of the SATCON technique for the various TC categories}
\begin{figure}
    \centering
    \includegraphics[width = 1\textwidth]{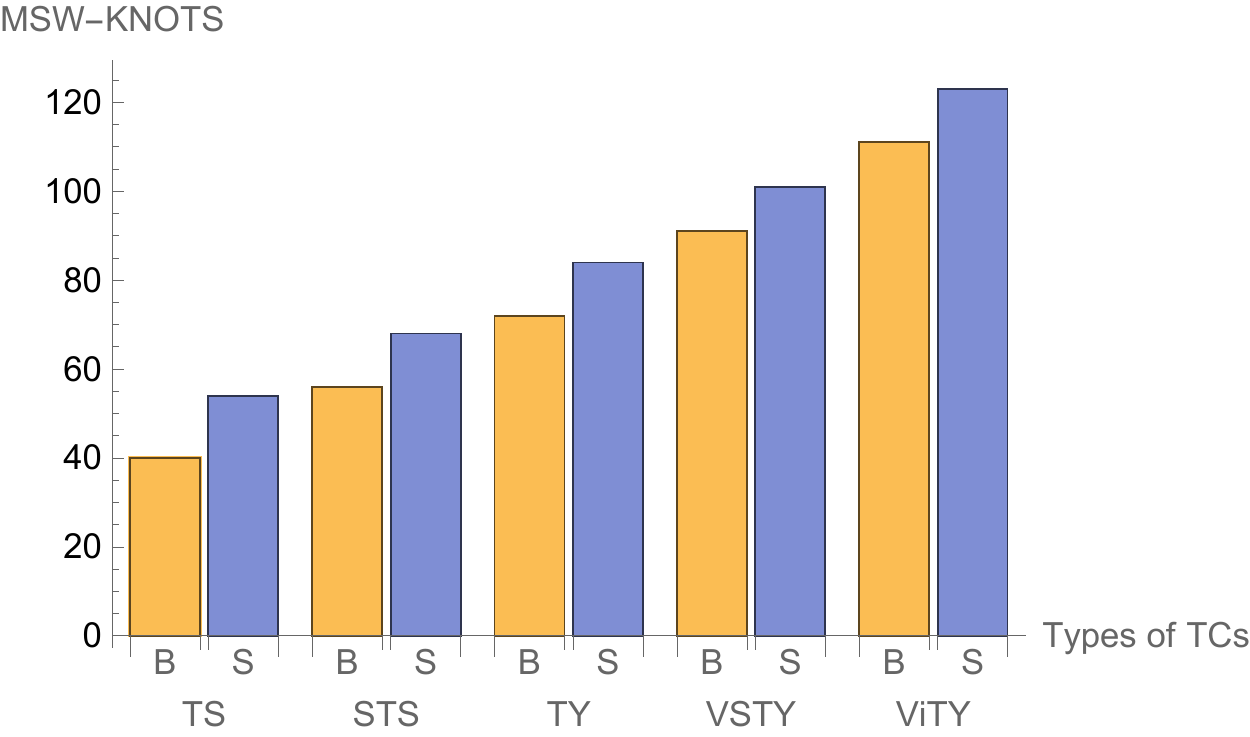}
    
    \caption{Compared the intensity estimation by SATCON and RSMC provided data for all TCs(stahe wise) over the west pacific basin during the year 2017-2021. 
    	B stands for RSMC provided data, S stands for SATCON algorithm data, ViTY stands for Violent Typhoon, VSTY for Very Strong Typhoon, TY for Typhoon, STS for Severe Tropical Storm, and TS for Tropical Storm}
    \label{fig2}
\end{figure}

The SATCON method and the intensity estimation data provided from the RSMC, Tokyo were used to analyse the average characteristics of tropical storms to violent typhoons over the west pacific basin between the year 2017-2021 in terms of the MSW (knots) and MSLP (hPa). It demonstrates that as TC goes to a higher category, the bias steadily reduces, being around 10-8 knots for a tropical storm to a severe tropical storm and 8-4 knots for a severe tropical storm to very strong typhoon,and violent typhoon.
Due to the small sample size, the bias value for the violent typhoon category, 12.21 kts, may not be indicative (table \ref{tab4}). Accordingly, the bias is reduced for stronger TCs, with the exception of violent typhoons, which is consistent with \cite{bib1,bib2} findings. Although the MAD is for typhoons, severe typhoons, and tropical storms, approximately 9-12 kts, and approximately 7-11 kts for very strong and violent typhoons (figure \ref{fig2}). For all TC categories, the overestimation is statistically significant at a $99\%$ level of confidence.

\begin{sidewaystable}
\sidewaystablefn%
\begin{center}
\begin{minipage}{\textheight}
\caption{Compared the SATCON and RSMC, Tokyo provided data (in terms of MSW and MSLP) of TCs (stage wise) over the west pacific basin during the year 2017-2021}\label{tab4}
\begin{tabular*}{\textheight}{@{\extracolsep{\fill}}lccccccccc@{\extracolsep{\fill}}}
\toprule%
&&Best &Best &&&&Mean&\\
 Category  & Total no. & track	&  track  & SATCON  & SATCON & BIAS  &   absolute & RMSD \\
  & of cases & range  &  MSW (A\footnotemark[1])& range & MSW (B\footnotemark[1])  & (A-B) & difference &  \\
 
\midrule
Violent Typhoon&34&105-130&110.8&116-144&123.01&-12.21&7.43&9.24 \\
Very Strong Typhoon&79&85-105&90.89&73-138&101.36&-10.47&10.97&12.5 \\
Typhoon&167&64-85&72.23&62-100&84.07&-11.84&9.61&11.43 \\
Severe Tropical Storm&108&48-64&55.73&46-92&68.21&-12.48&11.06&13.74 \\
Tropical Storm&251&34-48&39.59&40-74&53.93&-14.34&10.89& 12.96\\
\botrule
 MSLP (hPa)\\
\toprule
&&Best &Best &&&&Mean&\\
 Category  & Total no. & track	&  track  & SATCON  & SATCON & BIAS  &   absolute & RMSD \\
  & of cases & range  &  MSLP (A\footnotemark[1])& range & MSLP (B\footnotemark[1])  & (A-B) & difference &  \\
  \midrule
Violent Typhoon&28&920-928&925.56&922-940&928.45&-2.89&6.94&7.24 \\
Very Strong Typhoon&75&932-966&962.63&930-972&957.17&5.46&8.41&9.58 \\
Typhoon&168&964-984&981.81&953-983&975.73&6.08&9.08& 10.43\\
Severe Tropical Storm&106&983-990&980.58&965-990&976.3&4.28&7.83&9.12 \\
Tropical Storm&267&983-1002&1003.29&984-1004&1002.03&1.26&6.14& 7.01\\
 \midrule
\end{tabular*}
\footnotetext[1]{\label{1}In each range there are a number of cases where the mean has been calculated and represented by A and B. \\ In each range of the Best Track data, A represents the mean values of the number of cases. \\ In each range of the SATCON data, B represents the mean values of the number of cases.}
\end{minipage}
\end{center}
\end{sidewaystable}

The MAD value for typhoons and intense typhoons (8-10 kts) is consistent with \cite{bib1,bib2}'s prior findings. A slight increase in MAD values of intensity over the west pacific basin is recorded for tropical storm, severe tropical storm and very strong typhoon compared with \cite{bib1,bib2} conclusion. For tropical storms, severe tropical storms, typhoons, and very strong typhoons, the RMSD values over the west Pacific are approximately 11-14, and for violent typhoons, they are less than 10 kts. The Violent typhoon RMSD estimates across the west pacific ($<$10 kts) are consistent with \cite{bib1,bib2} earlier observations.
However, compared to \cite{bib1,bib2} findings, the RMSD values over the west pacific for the tropical storm, severe tropical storm, typhoon and very strong typhoon category are marginally greater (11-14 kts).

As a result, the intensity is understated (negative bias) in terms of MSLP by about -3 hPa for violent typhoons and 1-6 hPa for all other types of storms. For tropical storms, the MAD is roughly 6 hPa, and for all other storm types, it is between 7-9 hPa. All storm types have RMSD values between 7 and 11 hPa.

The SATCON algorithm shows the overestimation of the intensity of TCs during the begining stage of formation and up to T2.5, it may be seen from this. But after that, it is discovered that its performance is fairly good in measuring the intensity of stronger TCs (more than Severe Tropical storm). In the SATCON method, creates a single estimate from several TC intensity estimations derived from objective intensity algorithms. The major component of the SATCON model, ADT 9.0, feeds continuous inputs into the model every 30 minutes, whilst the microwave sounder satellite feeds irregular intensity inputs into the model, which are then extrapolated to hourly estimations. The final SATCON estimate is produced by combining these interpolated estimates with ADT estimates. An objective method evolved from the original Dvorak Technique is used by the ADT to calculate intensity. Up to T2.5 in the first development phase, the cloud organisation pattern is not clearly specified. At this time, the Dvorak Technique is unable to comprehend the intricate details of cloud patterns. Because of this, both the ADT 9.0 technique and SATCON overstate the intensity estimations based on the methodology's pre-defined fixed cloud pattern, primarily the central dense overcast (CDO) and eye pattern, regardless of whether it is a curved band or shear pattern. It goes without saying that the shear pattern TCs have a maximum strength of T3.0 and that majority of the TCs over the west pacific originate from shear patterns under the influence of monsoon circulation. As a result, when the intensity is T2.0 or higher, the ADT 9.0 version and SATCON are utilised globally. Additionally, the SATCON algorithm is reasonable good for T3.0 and more because for TCs whose intensity is more than T3.0, they show clear cloud pattern i.e. either eye pattern or CDO. In addition, SATCON used the new ADT 9.0 methodology, which integrates infrared sensor, short-wave infrared imaging sensor, visible imaging sensor, and microwave images to find phenomena that the original Dvorak Technique was unable to find, such as secondary eye-wall formation, double eyewall structure, the centre in the presence of cirrus canopy, coiling of convective clouds (in the presence of cirrus) around the centre, and eye-wall replacement cycle \cite{bib21}.

Given the foregoing, forecasters can utilize the SATCON technique to estimate intensity in the case of stronger TCs (T3.0 or more).
As cloud organization patterns are not clearly defined in the beginning stage, and the automated approach of ADT (a attribute of SATCON) selects pre-established patterns, overestimating of the intensity results, it is not suitable to cyclogenesis and the begining phase of TC formation.
Forecasters can, however, accurately estimate TC intensities based on SATCON data by using the bias, RMSD, and MAD calculated in this study.
\subsection{ Capability of SATCON algorithm in various seasons}
\subsubsection{The pre-monsoon season's capabilities of the SATCON algorithm}
\begin{figure}
	\centering
	\includegraphics[width=1\textwidth]{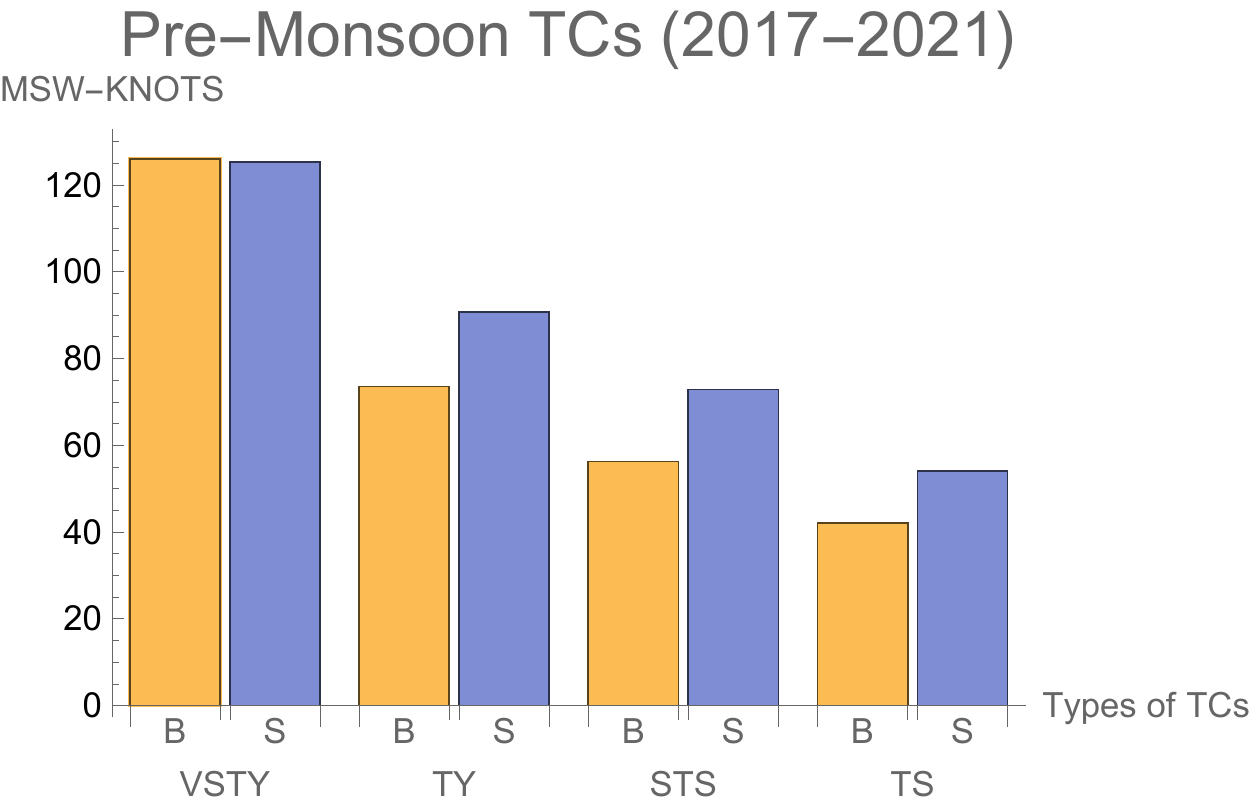}
	\caption{Compared the intensity estimation by SATCON and RSMC provided data for pre-monsoon season. 
		B stands for RSMC provided data, S stands for SATCON algorithm data, VSTY for Very Strong Typhoon, TY for Typhoon, STS for Severe Tropical Storm, and TS for Tropical Storm}
	\label{fig3}
\end{figure}

SATCON TC MSW (kts) and MSLP (hPa) estimates' capability in comparison to RSMC Tokyo intensity estimate data for TCs developed over the west Pacific during the pre-monsoon are shown in tables \ref{tab5} and \ref{tab6}. With the exception of very strong typhoons, the bias value stays high during all phases of TCs at roughly 11-18 kts. A very strong typhoon has a bias value of less than 1 kts (figure \ref{fig3}). The smaller sample size may be the cause of the very strong typhoon's unrepresentative value. According to the student's t-test for all types of TCs, the difference is significant at a $99\%$ level of confidence. 

For tropical storms, severe tropical storms, and typhoons, the MAD is approximately 12-19 kts, and for very strong typhoons, it is approximately 4 kts. The RMSD ranges between 13 and 19 kts for tropical storms, severe tropical storms, and typhoons, and between 4 and 5 kts for very strong typhoons. This runs counter to \cite{bib1,bib2} past findings.The average SATCON intensity, in turn, overestimates the MSW in the pre-monsoon season by around 11 kts and underestimates the average MSLP estimations by nearly 7 hPa, as demonstrated in Table \ref{tab5}.

\subsubsection{The post-monsoon season's capabilities of the SATCON algorithm}
\begin{figure}
    \centering
    \includegraphics[width = 1\textwidth]{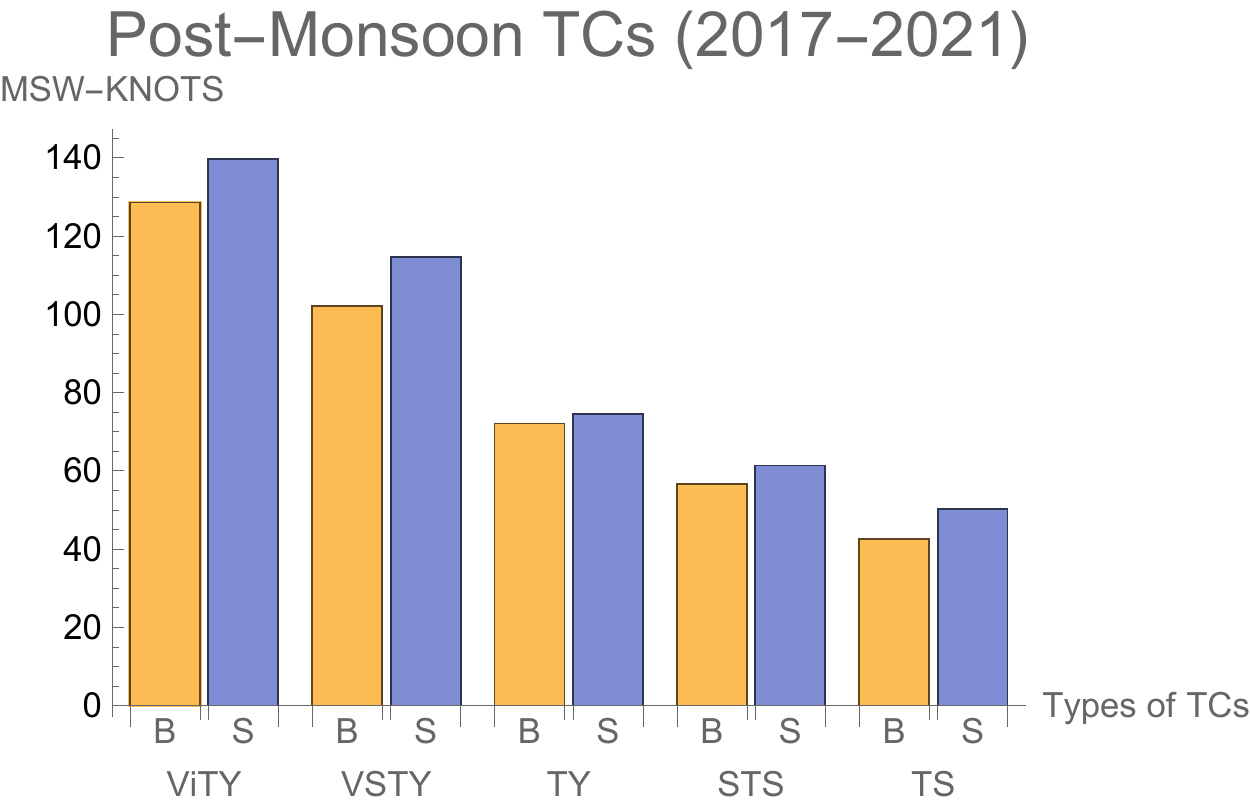}
    \caption{Compared the intensity estimation by SATCON and RSMC provided data for post-monsoon season. 
    B stands for RSMC provided data, S stands for SATCON algorithm data, ViTY stands for Violent Typhoon, VSTY for Very Strong Typhoon, TY for Typhoon, STS for Severe Tropical Storm, and TS for Tropical Storm}
    \label{fig4}
\end{figure}

Tables \ref{tab5} and \ref{tab6} show the capability of SATCON's algorithm of TCs MSW (kts) and MSLP (hPa) estimations compared to RSMC, Tokyo provided data of intensity estimates for TCs developed across the west Pacific during the year 2017-2021's post-monsoon. When the strength rises, the bias steadily decreases between 2 and 7 knots for tropical storms, severe typhoons, typhoons, and between 11 to 13 knots for extremely strong and violent typhoons (figure \ref{fig4}). The student's t-test for all forms of TC indicates that the difference is significant at a $99\%$ level of confidence. The bias value for the typhoon stage is consistent with \cite{bib1,bib2} past research.

The MAD for a tropical storm is approximately 11 knots, for TC categories such as a severe tropical storm, typhoon, very strong typhoon, and for violent typhoon, it is between 8 and 10 knots. The results of \cite{bib1,bib2} are supported by the MAD values throughout the west Pacific for severe tropical storms, typhoons, very strong typhoons, and violent typhoon stage (8-10 kts).
However, compared to \cite{bib1,bib2} findings, the MAD values for the tropical storm stage are a little bit higher (11 kts). For all storm types, the RMSD across the west Pacific is approximately 10-14 kts.
This runs counter to \cite{bib1,bib2} earlier research conclusions. Table \ref{tab5} demonstrates that the average SATCON intensity underestimates the average MSLP estimates by around 5 hPa while overestimating the average MSW during the post-monsoon season by nearly 9 kts.
\begin{sidewaystable}
\sidewaystablefn%
\begin{center}
\begin{minipage}{\textheight}
\caption{Compared the SATCON and RSMC, Tokyo provided data (MSW/MSLP) for TCs over the west pacific basin during the year 2017-2021 as pre-monsoon, post-monsoon, and annual }\label{tab5}
\begin{tabular*}{\textheight}{@{\extracolsep{\fill}}lccccccccc@{\extracolsep{\fill}}}
\toprule%

 Season  & Total no. 	& Best track    & SATCON & BIAS  &  Mean absolute & RMSD \\
  & of cases   &  MSW (A\footnotemark[1])& MSW (B\footnotemark[1])  & (A-B) & difference &  \\
 Pre-Monsoon&158&64.85&75.59&-10.74&11.53&13.96 \\
  Post-Monsoon&308&59.51&66.26&-6.75&9.38&12.21 \\
  Annual Season&466&62.18&70.92&-8.74&10.83&13.07 \\
\midrule

\botrule
 MSLP (hPa)\\
\toprule
 Season  & Total no. 	& Best track    & SATCON & BIAS  &  Mean absolute & RMSD \\
  & of cases   &  MSW (A\footnotemark[1])& MSW (B\footnotemark[1])  & (A-B) & difference &  \\
  Pre-Monsoon&156&984.13&976.3&7.83&8.55& 10.59\\
  Post-Monsoon&359&959.39&958.16 &1.23&5.34&7.24\\
  Annual Season&515&971.76&967.23&4.53&6.86&8.79 \\
 \midrule
 
\end{tabular*}
\end{minipage}
\end{center}
\end{sidewaystable}
\begin{table}[]
    \centering
    \begin{tabular}{c c c c c c c c c }
    \hline
         & & Best & Best &  &  &  & Mean &  \\
       Category &Season&Track& Track &SATCON&SATCON&BIAS&Absolute&RMSD\\
        of TC && interval &MSW & interval & MSW(B\footnotemark[1]) & (A-B) & difference& \\
        &&&(A\footnotemark[1]) \\
        \hline
        Violent &Pre-Mon&-&-&-&-&-&-&- \\
        Typhoon &Post-Mon& 105-130&128.57&120-144&139.73&-11.16&8.68&13.34\\
        \hline
        Very & Pre-Mon&80-110&126.03&90-120&125.30&0.73&4.26&4.92\\
        strong & Post-Mon&85-105&102.08&112-132&114.57&-12.49&9.52&14.19\\
        Typhoon & \\
        \hline
        Typhoon & Pre-Mon&64-85&73.58&79-100&90.77&-17.19&18.44&18.24 \\
        &Post-Mon&65-90&72.12&72-98&74.46&-2.34&8.24&10.18 \\
        \hline
        Severe& Pre-Mon&48-64&56.25&69-81&72.87&-16.62&16.54&17.08 \\
        Tropical&Post-Mon&44-63&56.66&57-87&61.36&-4.69&9.24&12.47\\
        Storm \\
        \hline
        Tropical& Pre-Mon&34-48&42.01&16-70&54.11&-12.1&12.19&13.48 \\
        Storm &Post-Mon&30-50&42.62&39-74&50.19&-6.92&10.58&12.84\\
        \hline
       
    \end{tabular}
    \caption{Compared the SATCON and RSMC, Tokyo provided data for TCs (stage wise) over the west pacific basin as pre-monsoon and post-monsoon during the year 2017-2021}
    \label{tab6}
    \footnotetext[1]{In each range there are a number of cases where the mean has been calculated and represented by A and B. In each range of the Best Track data, A represents the mean values of the number of cases. In each range of the SATCON data, B represents the mean values of the number of cases.}
\end{table}

\section{Conclusion} \label{sec4}
The key takeaways from the results and discussions above are listed below.

Tropical storm, severe tropical storm, typhoon, very strong typhoon, and violent typhoon types of TCs were examing in terms of intensity (`T' number) estimates across the west pacific basin from 2017 to 2021 using data from RSMC, Tokyo and the SATCON algorithm. As TCs progress through the initial phase of development, the range of overestimation of SATCON intensity estimation decreases. The result for T6.0-T7.0 may not be representative due to the sample size.

When we compared the SATCON algorithm's output with the data provided by the RSMC, we found that during the pre-monsoon, the SATCON algorithm overestimated tropical storms by about 13 kts, severe tropical storms by about 17 kts, and typhoons by about 19 kts. During the post-monsoon, the SATCON algorithm overestimated tropical storm by about 11 kts, and severe tropical storm, typhoon, very strong typhoon and violent typhoon by about 9kts.

We demonstrate that SATCON is more effective in the post-monsoon across the west pacific basin than in the pre-monsoon by comparing the algorithm results.
\section*{Acknowledgements}
The RSMC Tokyo and CIMSS-SATCON are thanked by the authors for providing the information used in this article. The authors appreciate the anonymous peer reviewers' insightful criticism, which helped the paper's quality.
\section*{Author Statement}
Monu Yadav: Conceptualization, investigation, data curation, methodology, validation, preparation of tables/figures. Laxminarayan Das: Supervision, reviewing and editing.

\bibliography{sn-bibliography}


\end{document}